\documentclass[twocolumn,showpacs,preprintnumbers,amsmath,amssymb]{revtex4-1}
\usepackage[dvipdfmx]{graphicx}
\usepackage{dcolumn}% Align table columns on decimal point
\usepackage{bm,color}% bold math

\newcommand{\be}{\begin{eqnarray}}
\newcommand{\ee}{\end{eqnarray}}
\newcommand{\nn}{\nonumber\\}

\begin{document}

\title{Phase diagrams of the extended Bose-Hubbard model in one dimension
by Monte-Carlo simulation with the help of a stochastic-series expansion}
% Force line breaks with \\

\author{Keima Kawaki, Yoshihito Kuno, and Ikuo Ichinose}
\affiliation{
Department of Applied Physics, Nagoya Institute of Technology, 
Nagoya, 466-8555, Japan}

\date{\today}% It is always \today, today,

\begin{abstract}
In this paper, we study phase diagrams of the extended Bose-Hubbard 
model (EBHM) in one dimension by means of the quantum Monte-Carlo (QMC) 
simulation using the stochastic-series expansion (SSE).
In the EBHM, there exists a nearest-neighbor repulsion as well as the on-site
repulsion.
In the SSE-QMC simulation, the highest particle number at each site, $n_c$, is also
a controllable parameter, and we found that the phase diagrams depend
on the value of $n_c$.
It is shown that in addition to the Mott insulator, superfluid, density wave, 
the phase so-called Haldane insulator and supersolid appear in the phase diagrams, 
and their locations in the phase diagrams are clarified.
\end{abstract}

\maketitle
%%%%%%%%%%%%%%%%%%%%%%%%%%%%%%%%%%%%%%%%
\section{Introduction}

Quantum many-body systems in one spatial dimension (1D) have
strong fluctuations compared with higher-dimensional systems, and as a result,
they sometimes have exotic quantum phases and nontrivial phase diagrams 
that cannot be obtained by mean-field theories. 
Recently, experiments on ultracold atomic systems can produce controllable 
and versatile strongly-correlated systems on an optical lattice \cite{coldatoms}.
There, the strong correlations mean large on-site and off-site atomic interactions \cite{Dutta}, 
a strong artificial magnetic field \cite{Goldman}, geometrical frustrations, 
e.g., on triangular and honeycomb lattices \cite{Jotzu}, etc. 
In this paper, the phase diagram of an extended Bose-Hubbard model (EBHM) 
on the 1D lattice is investigated by means of one of the most
reliable numerical methods, i.e., the quantum Monte-Carlo (QMC) simulation with 
the stochastic-series expansion (SSE) \cite{SSE}.

This model is expected to have a rich phase diagram due to large fluctuations and 
nearest-neighbor interactions.
It is believed that the model has similar properties of spin chain models, 
which are important models in condensed matter physics. 
The previous studies \cite{Altman} discussed that in the case of the strong
on-site interaction, the particle number at each site is restricted to be less than two, 
and as a result, the three body constraint, $(a^{\dagger})^{3}=0$, seems to appear
\cite{Daley,Diehl,Diehl2,Chen}.
Under this constraint, the EBHM can be mapped to a spin-1 XXZ-type model 
by using the Holstein-Primakoff transformation \cite{Altman,Berg}. 
From this relationship between the EBHM and quantum spin model,
one may expect the existence of an interesting phase, i.e.,  
Haldane insulator (HI), which is similar to the Haldane phase in the 
quantum spin system \cite{Affleck,Kennedy}. 
So far, a number of the numerical studies \cite{1DDMRG,1DDMRG2,1DDMRG3} 
investigated the phase diagram of the EBHM in the canonical ensemble
incorporating the constraint $(a^{\dagger})^{3}=0$.
In most of these studies, the filling fraction is fixed to unity, although some of 
them studied other low-filling cases. 
Furthermore, we expect that real experimental set up may relax such 
a three body constraint, then the mapping of the EBHM to the spin-1 model is 
not necessarily applicable. 
Therefore, the EBHM may have a richer phase diagram than the spin model. 
In particular, the detailed phase diagram of the EBHM in the 
grand-canonical ensemble is not completely understood yet.

In this paper, we consider the grand-canonical ensemble of the EBHM and study
the phase diagram by the SSE-QMC simulations.
In fact, the SSE-QMC simulation is suitable for the study on
the grand-canonical ensemble as 
large system-size calculation is possible due to less memory consumption 
compared to other numerical methods, e.g., the exact diagonalization method.
Obtained phase diagram exhibits various phases with various filling fractions.
For example, the aforementioned HI appears not only at the unit filling but also at the
half filling.

The paper is organized as follows.
In Sec.II, we introduce the EBHM and explain the SSE-QMC simulation.
Various quantities to identify phases are introduced.
In Sec.III, results of the numerical study are presented.
In the practical simulation, the maximum number of particles at
each site ($n_c$) and also the value of the next-nearest-neighbor repulsion ($V$)
are fixed.
Phase diagrams in the [on-site repulsion]-[chemical potential 
(i.e., average particle number)] are obtained.
Results show the dependence of the phase diagrams on the value of $n_c$.
System-size dependence of the results are also carefully examined.
Section IV is devoted for discussion and conclusion.
 
%%%%%%%%%%%%%%%%%%%%%%%%%%%%%%%%%%%%%%%%%%%%%%%%%%%%%%%%%%%%%%%%%
\section{Extended Bose-Hubbard model and quantum MC simulation with SSE}

We start with the EBHM defined on a 1D lattice whose
Hamiltonian $H_{\rm EBH}$ is given as
\begin{eqnarray}
H_{\rm EBH}&=&\sum_a \biggl[-J(\hat{\psi}^{\dagger}_a\hat{\psi}_{a+1}
+\hat{\psi}^{\dagger}_{a+1}\hat{\psi}_a)  
+{U\over 2}(\hat{\rho}_{a}-1)\hat{\rho}_a\nn
&&+V\hat{\rho}_{a} \hat{\rho}_{a+1} \biggr], \nn
\hat{\rho}_a&\equiv& \hat{\psi}^{\dagger}_a\hat{\psi}_a,
%+\nu\sum_{i}\rho_{i},
\label{BHM}
\end{eqnarray}
where $\hat{\psi}^{\dagger}_a$ and $\hat{\psi}_a$ 
are creation and annihilation operators of 
boson at site $a$, respectively, and $\hat{\rho}_a$ is the number operator.
The coefficient $J$ represents the hopping strength, $U$ is the on-site interaction, 
and $V(>0)$ is the nearest-neighbor (NN) replusive interaction generated by, e.g., 
a dipole-dipole interaction in gases loaded on the optical lattice \cite{1Doptical_DDI,DDI}. 
In the cold atomic gas system, the above on-site repulsion $U ( > 0)$ represents 
the sum of $s$-wave scattering interaction
$U_{s}$ and on-site dipole-dipole interaction $U_{d}$; $U= U_{s}+U_{d}$. 
The $s$-wave scattering amplitude $U_{s}$ is highly controllable by the 
Feshbach resonance \cite{Feshbach}. 
In practical experiments, the ratio $V/U$ is highly controllable 
by using the combination of the Feshbach resonance and selection of spices of
loaded atoms \cite{Dutta,Baier}.

The global phase diagram of the EBHM in Eq.(\ref{BHM}) is important and
we shall clarify the low-filling phase diagram of the EBHM
by means of the most reliable numerical method, i.e., 
the SSE-QMC simulation \cite{SSE}.  
In the SSE-QMC simulation, the partition function is expanded as 
\begin{eqnarray}
Z_{\rm EBH}&=&\mbox{Tr}(e^{-\beta (H_{\rm EBH}-\mu N)})  \nonumber  \\
&=&\sum_{n=0}^\infty
{1 \over n!}\mbox{Tr}\Big(-\beta (H_{\rm EBH}-\mu N)\Big)^n,
\label{ZSSE}
\end{eqnarray} 
where $\beta=1/(k_{\rm B} T)$, $k_{\rm B}$ is the Boltzmann constant,
$T$ is the temperature, $\mu$ is the chemical potential 
and $N=\sum_a\hat{\rho}_a$.
As we are interested in the ground-state phase diagram, we take 
$\beta\rightarrow$ large.
In the evaluation of $Z_{\rm EBH}$ in Eq.(\ref{ZSSE}), the particle-number eigen-states 
$\prod_a|\rho_a\rangle$, 
$\hat{\psi}^\dagger_a\hat{\psi}_a|\rho_a\rangle=\rho_a|\rho_a\rangle$,
are employed as a basis of quantum states. 
Then, the Hamiltonian $H_{\rm EBH}$ is divided into the diagonal part
(the $U$ and $V$-terms) and off-diagonal part (the $J$-term), 
and Eq.(\ref{ZSSE}) is re-expanded in powers of these parts.
Weight of each term in the expansion is determined by the MC methods.
The trace  in Eq.(\ref{ZSSE}) can be calculated by putting intermediate states 
between the Suzuki-Trotter decomposed Hamiltonians.
Here, Monte-Carlo sampling is applied for each decomposed Hamiltonian operator. 
In the sampling, the loop algorithm \cite{SSE} allows to create closed loops of transition
states along imaginary time (temperature) direction.

In this work, we consider the case of low fillings, and restrict the Hilbert space
$\{|\rho_a\rangle\}$ to $\rho_a=0,\cdots, n_c$ in evaluating $Z_{\rm EBH}$ in 
Eq.(\ref{ZSSE}), where $n_c$ is the largest particle number at each site.
In the practical calculation, we first concentrate on the case $n_c=2$ and $3$,
and later on we show results in the case of higher $n_c$.
In Refs.\cite{1DDMRG,1DDMRG2,1DDMRG3}, the EBHM was studied mostly by 
the density-matrix renormalization group (DMRG).
There, the average particle number per site was fixed to unity, i.e., 
$\rho={1 \over L}\sum_a\langle{\rho}_a\rangle=1$, where $L$ is the system size,
and the phase diagram in the $(U$-$V)$ plane was obtained.
In the present study, on the other hand, we employ the grand-canonical ensemble and 
vary the chemical potential, i.e., the average particle density, to obtain
the phase diagrams, 
although we focus on the low-filling region like $0<\rho<3$ at first. 
As far as we know, the phase diagram of the EBHM in the grand-canonical 
ensemble is a new result.
By studying the EBHM in the grand-canonical ensemble, we found that the model
has different phase diagrams depending on the value of $n_c$.
For the case of the unit filling $\rho=1$ and $n_c=2$, the phase diagram of the EBHM
was obtained by the DMRG methods \cite{2DDMRG}.
As we explain later on, the obtained phase diagrams by the SSE-QMC simulation 
in the present study are in good agreement with
the phase diagram obtained in Ref.\cite{2DDMRG}.

In the practical calculation, we put $\hbar =1$, $J =1$ (as the unit of energy), 
and $\beta = 200$, which corresponds to a very low temperature case
\cite{SSE} and employ the periodic boundary condition.
We calculated the average particle density and also order parameters 
as varying the chemical potential $\mu$ for fixed values of $U$ and $V$.

Before going into the study on the EBHM, we investigated the phase diagram
of the standard Bose Hubbard model {\em without} the NN repulsion, i.e., the $V=0$ case.
The results support the accuracy of our numerical code because
the well-established phase diagram of the Bose-Hubbard model was reproduced 
quite accurately. 

%%%%%%%%%%%%%%%%%%%%%%%%%%%%%%%%%%
%Fig.1
\begin{figure}[t]
\centering
\includegraphics[width=6.5cm]{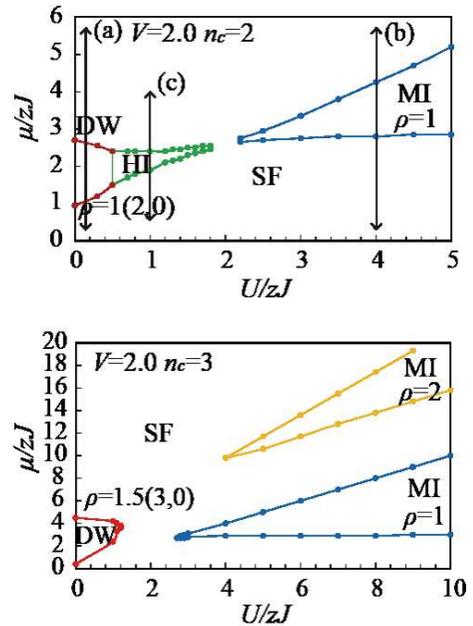}
\caption{(Color online) Phase diagram of the EBHM obtained by 
the SSE-QMC simulations for $V=2.0$ and $n_c=2 \ (n_c=3)$ in the upper (lower)
plane.
There are SF (superfluid), MIs (Mott insulators), DWs (density waves), and
HI  (Haldane insulator).
In the case of $n_c=2$ and at unit filling $\rho=1$, a direct transition from the MI to HI
does not takes place, instead, there is the tiny SF region.
$z$ is the number of the NN sites and in the present case $z=2$.
}
\label{QMCPD1}
\end{figure}
%%%%%%%%%%%%%%%%%%%%%%%%%%%%%%%%%%%%
%%%%%%%%%%%%%%%%%%%%%%%%%%%%%%%%%%
%Fig.2
\begin{figure}[t]
\centering
\includegraphics[width=8cm]{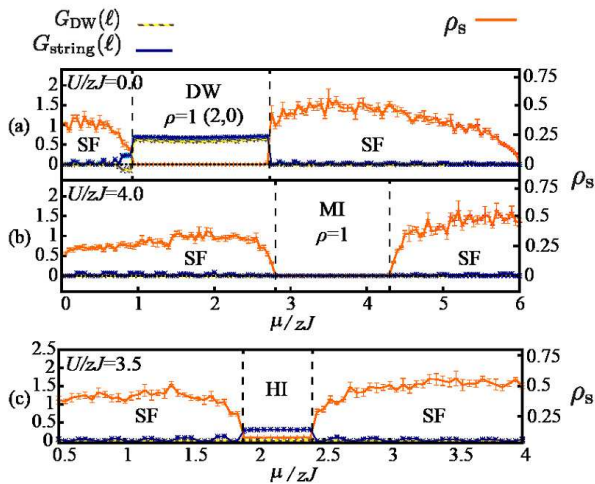}
\caption{(Color online) Various order parameters for 
$n_c=2$ and $V=2.0$ as a function of $\mu/zJ$.
(a) Results for $U/zJ=0.0$. (b) $U/zJ=4.0$. (c) $U/zJ=1$.
}
\label{QMCPD1corr}
\end{figure}
%%%%%%%%%%%%%%%%%%%%%%%%%%%%%%%%%%
%%%%%%%%%%%%%%%%%%%%%%%%%%%%%%%%%%
%Fig.3
\begin{figure}[t]
\centering
\includegraphics[width=7cm]{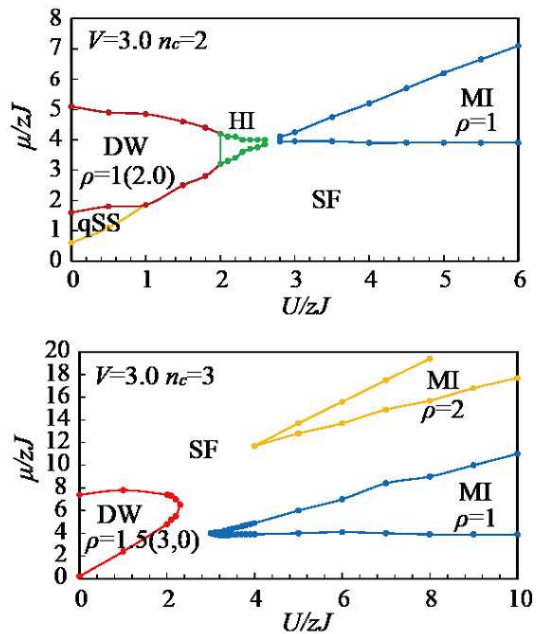}
\caption{(Color online) Phase diagram of the EBHM obtained by 
the SSE-QMC simulations for $V=3.0$ 
and $n_c=2 \ (n_c=3)$ in the upper (lower) plane.
There are SF, MIs, DWs, and HI as in the case of $V=2.0$.
In addition in the case $n_c=2$, there appear a phase that we call 
quasi-supersolid (qSS).
For details, see Fig.\ref{fig:dDW}.
As in the case of $V=2.0$ and $n_c=2$, the tiny SF region exists 
between the MI and HI
at unit filling $\rho=1$.}
\label{QMCPD2}
\end{figure}
%%%%%%%%%%%%%%%%%%%%%%%%%%%%%%%%%%

%%%%%%%%%%%%%%%%%%%%%%%%%%%%%%%%%%
%Fig.4
\begin{figure}[t]
\centering
\includegraphics[width=6.8cm]{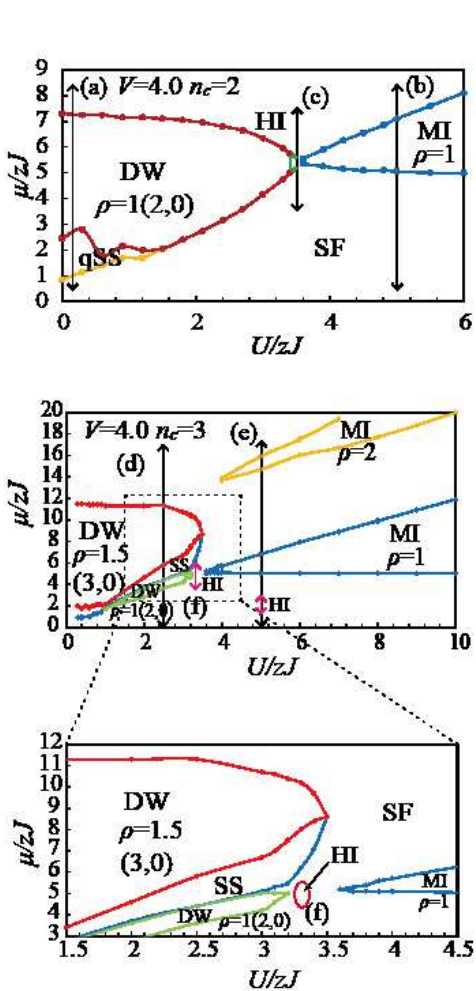}
\caption{(Color online) Phase diagram of the EBHM obtained by 
the SSE-QMC simulations for $V=4.0$ and $n_c=2~(n_c=3)$ 
in the upper (lower) plane.
There are SF, MIs, DWs, and HI.
qSS stands for the quasi-supersolid with a finite SF.
The phase diagram of $n_c=3$ is rather complicated
compared to the case of $n_c=2$.
In addition to the $\rho=1$ HI, there exists the $\rho={1 \over 2}$ HI, which
is discussed in Sec.IV.}
\label{QMCPD3}
\end{figure}
%%%%%%%%%%%%%%%%%%%%%%%%%%%%%%%%%%%%

To distinguish the phases, we measure various order parameters.
The superfluid (SF) order parameter $\rho_{\rm s}$ 
is related to the winding number 
of the boson world lines and defined as \cite{Pollock,Sandvik2}
\begin{equation}
\rho_{\rm s}={1 \over 2\beta L}\langle(N^+-N^-)^2\rangle,
\label{SF}
\end{equation}
where $N^{+} (N^-)$ is the total number of the hopping term
in the positive (negative) direction that appears in the MC simulation.
In the practical calculation of $\rho_{\rm s}$ in Eq.(\ref{SF}), 
we take the average of all 1D spatial configurations appearing in the 2D plain of
the 1D space and the expansion step of the completed loop.
For the 1D EBHM at large fillings, detailed path-integral MC simulations 
were performed in Ref.\cite{SForder}, and  
the Mott insulator (MI)$\leftrightarrow$SF phase transition
is observed as a Kosterlitz-Thouless transition.

Other order parameters that identify the density wave (DW) and 
the HI are the followings,
\begin{eqnarray}
&&G_{\rm DW}(\ell)=(-1)^\ell\langle \delta \rho_{a+\ell}\delta \rho_a\rangle, 
\label{OP} \\
&&G_{\rm string}(\ell)=\langle\delta \rho_{a+\ell}
e^{i\pi\sum_{a\leq k<a+\ell}\delta \rho_k}\delta \rho_a\rangle,
\label{OPs}
\end{eqnarray}
where $\delta \rho_a\equiv\rho_a-\rho$. 
$G_{\rm DW}(\ell)$ is a DW correlation function to detect the DW phase. 
On the other hand $G_{\rm string}(\ell)$ is a string-order correlation function, 
which can identify the HI phase. 
(The definition of this correlation function is slightly different from that used in 
the previous studies in Refs.\cite{1DDMRG,1DDMRG2,1DDMRG3}, i.e.,
in the definition $\delta \rho_a=\rho_a-\rho$, we do not fix 
the average density $\rho$ to unity as we employ the grand-canonical ensemble.)
A finite value of lim$_{\ell\rightarrow \infty}G_{\rm DW}(\ell)$ shows the existence
of the DW, which is expected to form for large $V$.
Finally, a finite value of lim$_{\ell\rightarrow \infty}G_{\rm string}(\ell)$ and the
{\em vanishing DW order} mean that the corresponding state is the HI.
This order is similar to the Haldane order in the anti-ferromagnetic (AF) spin chain,
and a typical configuration in the HI is shown in Ref.\cite{Berg}.
On the other hand, the nonvanishing  DW order always accompanies a finite string order.
%%%%%%%%%%%%%%%%%%%%%%%%%%%%%%%%%%
%Fig.5
\begin{figure}[t]
\centering
\includegraphics[width=8cm]{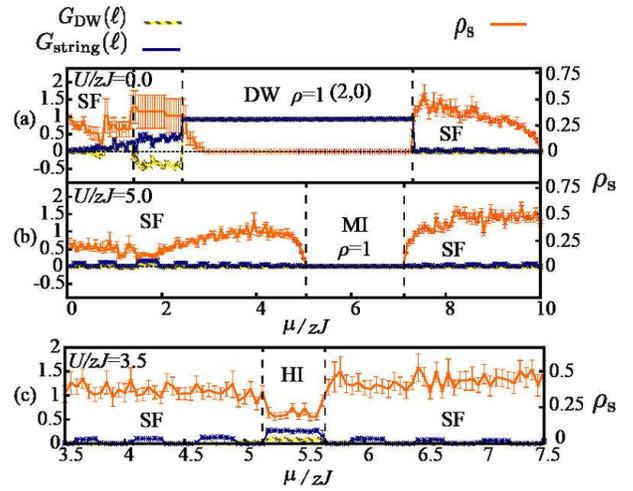}
\caption{(Color online) Typical correlation functions 
in the $n_c=2$ and $V=4.0$ phase diagram in Fig.\ref{QMCPD3}.
(a)$\sim$(c) correspond to  the lines indicated in Fig.\ref{QMCPD3},
respectively.
The system size is {\it L}$=32$. 
In calculating the order parameters Eqs.(\ref{OP}) and (\ref{OPs}), 
we used $\ell = L/2$.
SF (superfluid), MI (Mott insulator), DW (density wave), and
HI  (Haldane insulator).
}
\label{QMCPD2corr}
\end{figure}
%%%%%%%%%%%%%%%%%%%%%%%%%%%%%%%%%%%%

%%%%%%%%%%%%%%%%%%%%%%%%%%%%%%%%%%%%%%%%%%%%%%%%%
\section{Numerical results}

%%%%%%%%%%%%%%%%%%%%%%%%%%%%%%%%%%
%Fig.6
\begin{figure}[th]
\centering
\includegraphics[width=7cm]{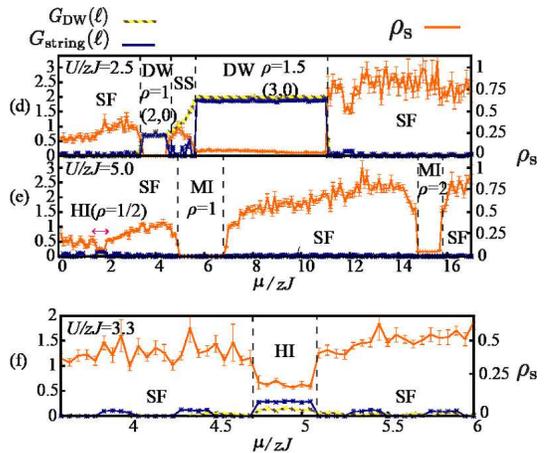}
\caption{(Color online) Typical correlation functions along the line 
(d)$\sim$(f) in the $V=4.0$ phase diagram in Fig.\ref{QMCPD3}.
The system size is {\it L}$=32$. 
In calculating the order parameters Eqs.(3)-(5), we used $\ell = L/2$.
SF (superfluid), MI (Mott insulator), DW (density wave), 
HI  (Haldane insulator) and SS (supersolid).
From the results in (e) and (f), we conclude the existence of the HIs
at $\rho={1 \over 2} \ (\mu/J\simeq 1.5)$ and 
$\rho=1 \ (\mu/J\simeq 4.7)$.
}
\label{QMCPD3corr}
\end{figure}
%%%%%%%%%%%%%%%%%%%%%%%%%%%%%%%%%%%%

\subsection{Phase diagrams for $n_c=2$ and $3$}

We first show the results of the $V=2.0$ case with the system size $L=32$.
Obtained phase diagrams are shown in Fig.\ref{QMCPD1} for the 
$n_c=2$ and $n_c=3$ cases.
There are four phases; the MI with $\rho=1~(n_c=2)$ and 
$\rho=1.5~(n_c=3)$, SF, DW and the HI.
For the $n_c=2$ case, the DW has the $|\cdots,2,0,2,0,\cdots\rangle$
configuration and the HI forms in relatively large-$U$ region,
in which a holon exists between every two doublons as indicated
by the finite $G_{\rm string}(\ell)$ \cite{Berg}.
It should be remarked that the $n_c=2$ EBHM is closely related to the 
spin-1 quantum Heisenberg spin chain \cite{Kennedy,Berg}.
The HI corresponds to the Haldane phase in the spin system. 
On the other hand for the case $n_c=3$, the DW with $\rho=1.5$ appears and 
the $|\cdots,3,0,3,0,\cdots\rangle$ configuration is realized there, whereas 
the HI does not form.
We have not found the DW with $\rho=1$ in the $(U/J-\mu/J)$ plane in the
present grand-canonical ensemble calculation, although we searched it in the 
low $\mu/J$ region.

%%%%%%%%%%%%%%%%%%%%%%%%%%%%%%%%%%
%Fig.7
\begin{figure}[t]
\centering
\includegraphics[width=8cm]{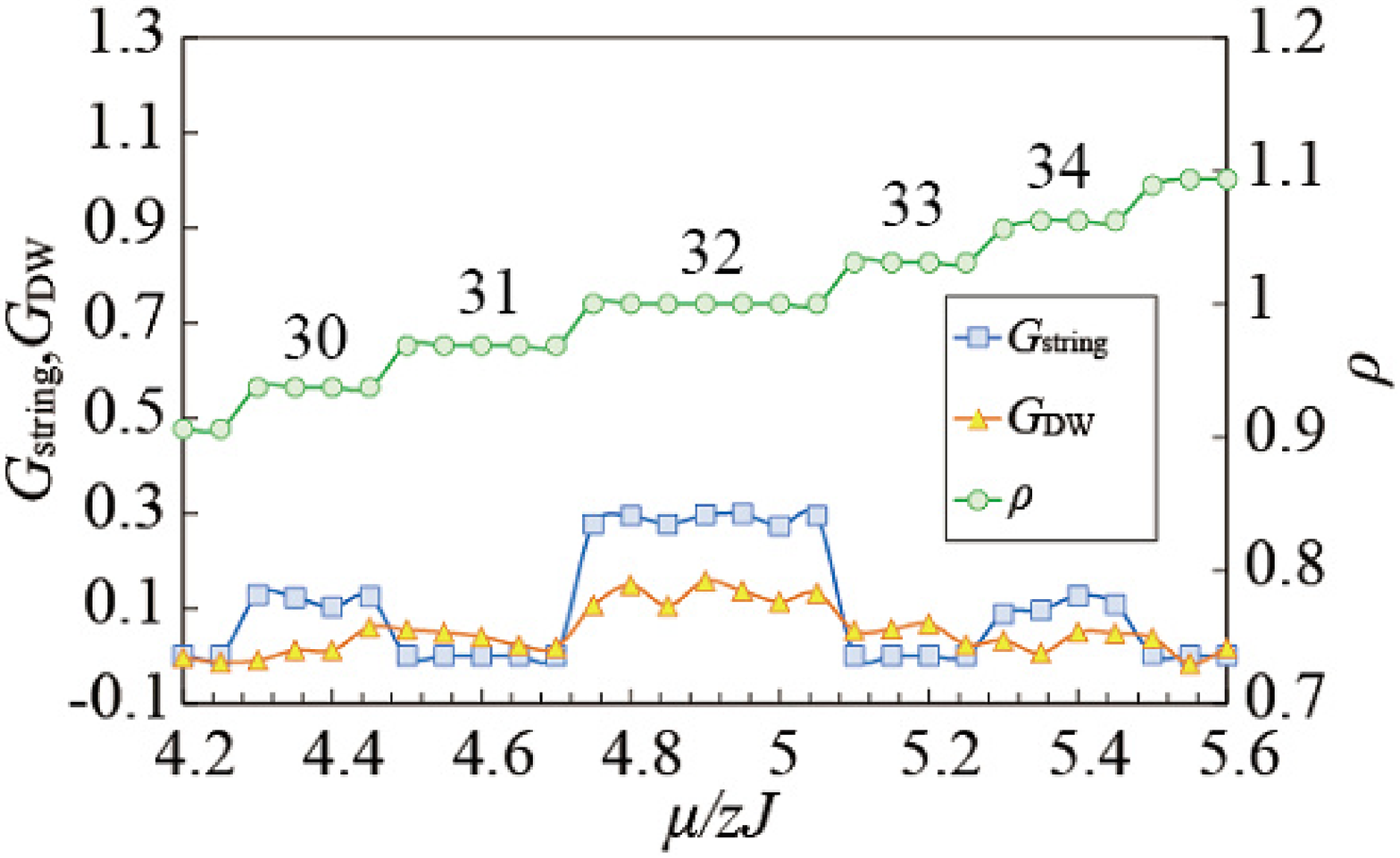}
\caption{(Color online) 
The string order, the DW and the density $\rho$ as a function of $\mu/zJ$
in the adjacent region of the $\rho=1$ HI.
$L=32$.
The numbers refer to total particle number in the system. 
Density $\rho$ exhibits a step-wise behavior synchronizing with the string order.
The result indicates that a state with a finite string order forms in the system
with an even-number of particle.
}
\label{evenodd}
\end{figure}
%%%%%%%%%%%%%%%%%%%%%%%%%%%%%%%%%%
%%%%%%%%%%%%%%%%%%%%%%%%%%%%%%%%%%
%Fig.dDW 8
\begin{figure}[t]
\centering
\includegraphics[width=7.5cm]{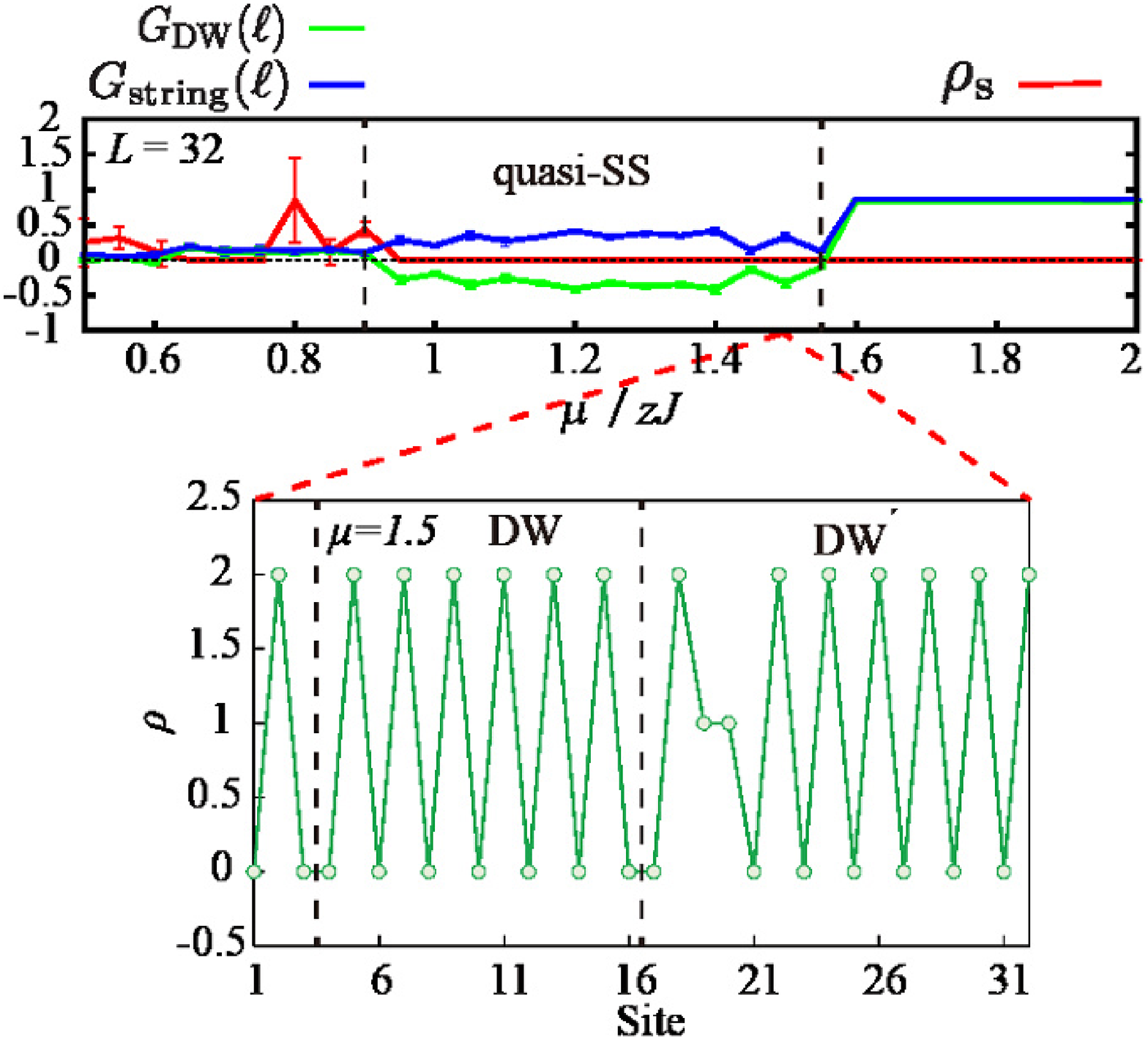}
\includegraphics[width=7.5cm]{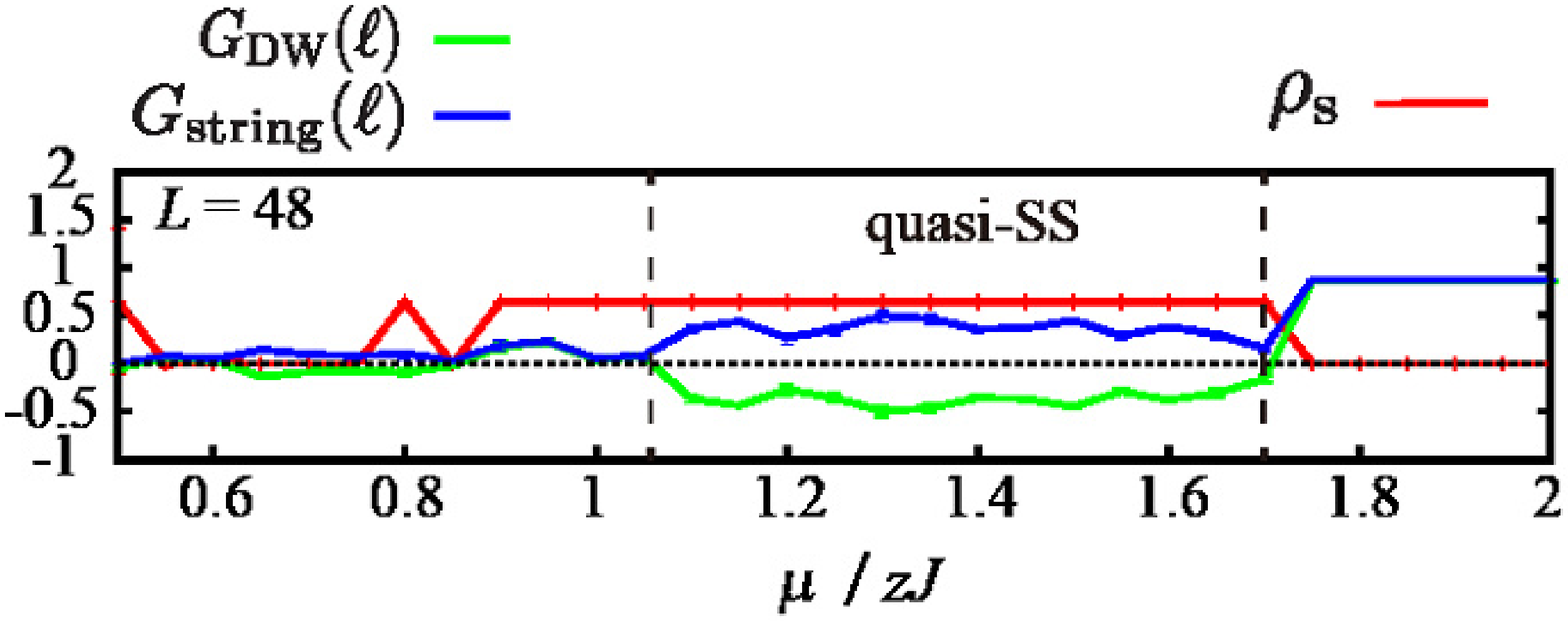}
\caption{(Color online) 
(Upper panel) Order parameters for the qSS in Fig.3 with $L=32$.
As the chemical potential decreases, the particle density decreases and 
extra holes and single occupied sites are generated in the $\rho=1$ DW 
existing for $\mu/zJ>1.6$.
Then the DW is divided into smaller DWs, and
the order parameter of the DW has negative values.
(Middle panel) A typical configuration generated by the MC simulation. 
(Bottom panel) Order parameters for $L=48$.
Walls dividing DW into smaller ones move
under the MC simulations in large system size.
As a result, the state has a finite SF density.
This is the reason why we call this state qSS.
}
\label{fig:dDW}
\end{figure}
%%%%%%%%%%%%%%%%%%%%%%%%%%%%%%%%%%

Typical behaviors of the order parameters are shown in Fig.\ref{QMCPD1corr}.
We also calculated the order parameters for the system sizes $L=40$ and $48$
and verified that the phase boundaries are stable.
In the MI and DW phases, the SF density is very low.
On the other hand, there exists a small but finite SF in the HI.
Later on, we shall show that the finite SF in the HI is a finite-size effect.
[More detailed analysis of the finite-seize effect will be given after showing 
the results of $V=4.0$.]

As Fig.\ref{QMCPD1corr} (c) shows,
the string order $G_{\rm string}(\ell)$ exhibits curious fluctuations in the SF 
that might stem from the relatively large density fluctuations, and these
fluctuations have small but finite spatial correlations. 
This unexpected behavior of  $G_{\rm string}(\ell)$ becomes clearer in the case of
$V=4.0$ that we shall study shortly.
We shall discuss the small but somewhat periodic regions with a finite 
$G_{\rm string}(\ell)$
after showing the phase diagrams of the $V=4.0$ case.

Next we show the phase diagram for $V=3.0$ in Fig.\ref{QMCPD2}.
Feature of the phase diagrams are almost the same with that in the case
of $V=2.0$, but the the region of the HI is getting smaller compared to the case
of $V=2.0$.
Furthermore in the phase diagram of $V=3.0$ with $n_c=2$,
in the vicinity of the DW, there exists a state that we call quasi-supersolid (qSS).
We shall discuss this state shortly.

Finally we show the phase diagram of the $V=4.0$ case in Fig.\ref{QMCPD3}.
In the case of $n_c=2$, there exists a small HI between the MI and DW
for $\rho=1$.
Behavior of the order parameters used to obtained the phase diagram
for $V=4.0$ with $n_c=2$ are shown in Fig.\ref{QMCPD2corr}.
On the other hand for $n_c=3$, the phase diagram is rather complicated,
i.e., the supersolid (SS) forms between two DWs with $\rho=1$ and $\rho=1.5$.
In the SS, a DW-like inhomogeneous state is realized,
and the average particle number is fractional $1<\rho<1.5$.
Particles (holes) move rather freely on the base of the $\rho=1$ ($\rho=1.5$) DW 
and as a result, the SF appears.
Interestingly enough, the phase diagram also indicates the existence of 
the $\rho={1 \over 2}$ HI as the order parameters in Fig.\ref{QMCPD3corr} show.
We shall discuss this HI in Sec.IV.

The calculations of the string order in Fig.\ref{QMCPD2corr} (c),
Fig.\ref{QMCPD3corr} (e) and (f) exhibit rather curious behavior.
It has a nonvanishing value in specific parameter regions of $\mu/zJ$,
which have a shell-like structure.
Figure \ref{evenodd} is a blow up of Fig.\ref{QMCPD3corr} (f). 
We show the density as a function of the chemical potential and find 
the step-wise behavior of the density synchronizing with the string order.
This result exhibits that a state with a finite string order forms in the system
with an even-number of particles
although in some regions no reduction of the SF is observed.
This might be a finite-size effect. See later discussion of the ``finite-size scaling"
analysis of the SF.
For example in the system with 34 particles, typical configurations are produced 
from those of the $\rho=1$ HI
by adding one doublon ($\rho_a=2$) to the system or replacing 
a singlton ($\rho_a=1$) with a triplon ($\rho_a=3$).

Let us briefly comment on the phase that we call qSS, which 
exists in the phase diagrams for $n_c=2$ in Figs.\ref{QMCPD2} and \ref{QMCPD3}.
As the chemical potential decreases, the particle density decreases from the 
$\rho=1$ DW. 
As a result of depletion of particles, the DW is divided into a few parts by 
``domain walls".
A typical configuration of the qSS obtained by the SSE-QMC simulation is shown 
in Fig.\ref{fig:dDW}.
A pair of holes plays a role of ``domain wall".
The SSE-QMC simulation shows that the $G_{\rm DW}$ has negative values.
As the system is getting large, pairs of holes (domain walls) are mobile, 
and as a result this state has a finite SF as seen in Fig.\ref{fig:dDW}.
In the large system size limit {\em with keeping the density of particle constant}, 
it is expected that the DW order parameter tends to vanish while a small but finite 
SF remains due to the mobility of domain walls.
This is the reason why we call that phase qSS.

%%%%%%%%%%%%%%%%%%%%%%%%%%%%%%%%%%
%Fig.9
\begin{figure}[t]
\centering
\includegraphics[width=6.5cm]{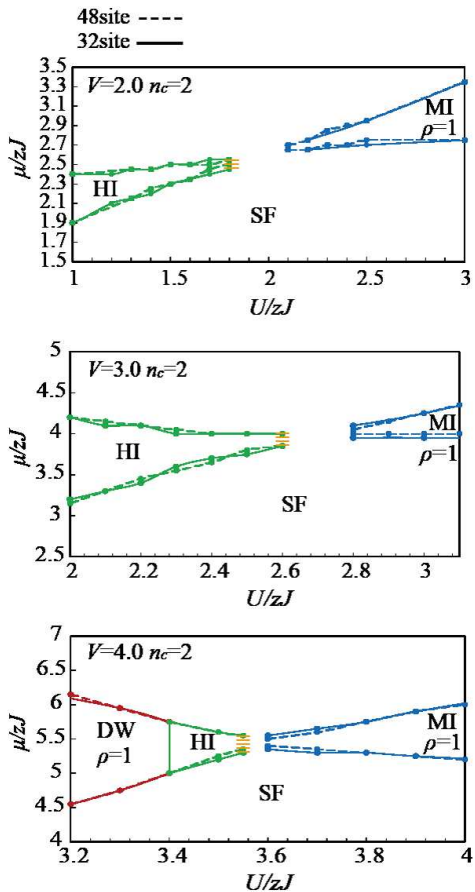}
\caption{(Color online) 
System-size dependence of the phase boundaries in the case of $n_c=2$. 
Simulations of the system sizes $L=32$ and $L=48$ exhibit almost the same
phase boundaries for all phase transitions.
This indicates that the system size $L=32$ reaches a scaling region of the 
thermodynamic limit. 
We have also verified other cases and obtained a similar system-size dependence.
}
\label{QMCPD_HI}
\end{figure}
%%%%%%%%%%%%%%%%%%%%%%%%%%%%%%%%%%%%

%%%%%%%%%%%%%%%%%%%%%%%%%%%%%%%%%%
%Fig.10
\begin{figure}[h]
\centering
\includegraphics[width=6.5cm]{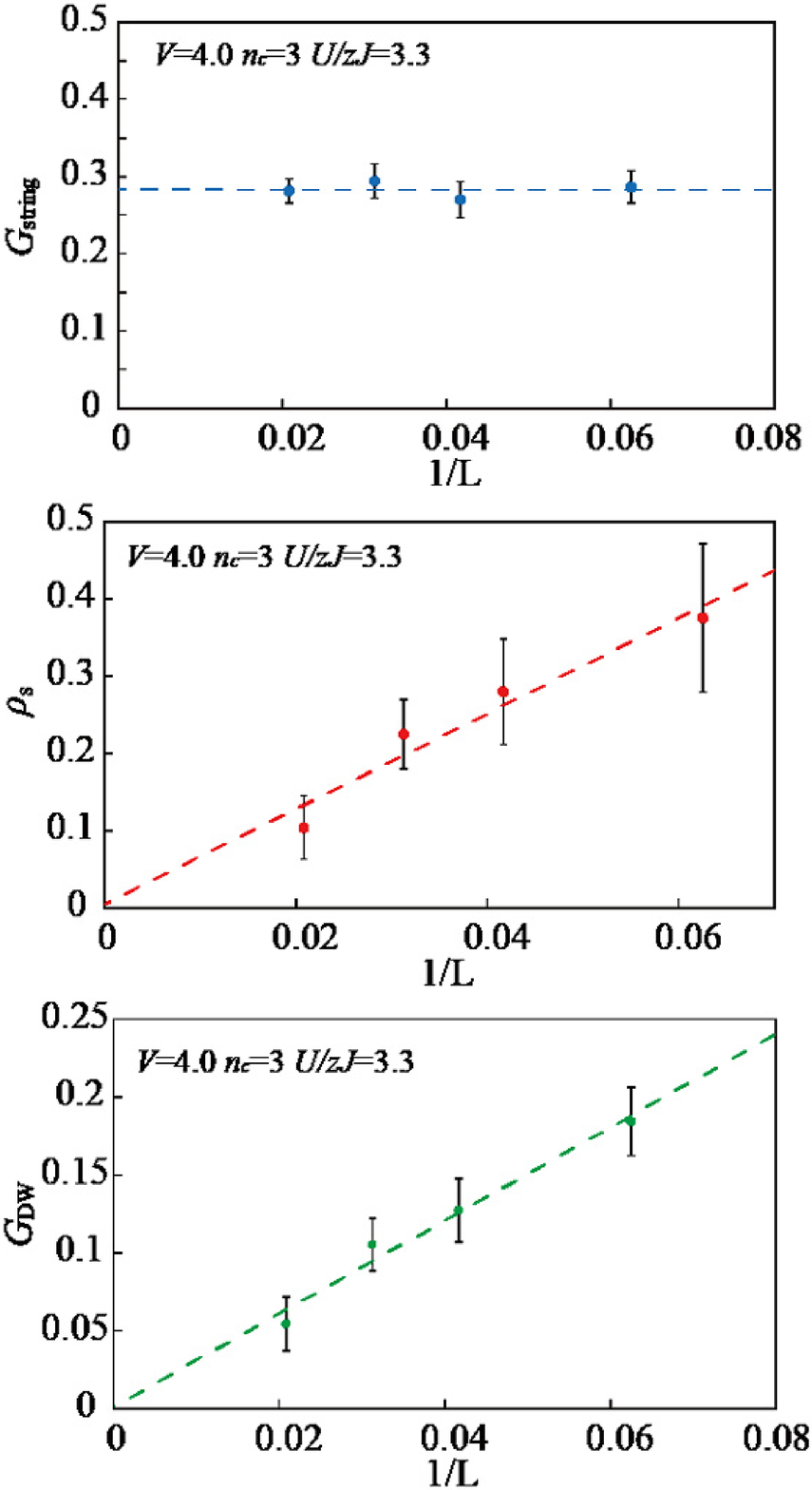}
\caption{(Color online) 
The string order (upper panel), SF (middle panel) and DW (bottom panel) 
in the $\rho=1$ HI as a function of $1/L$.
$G_{\rm string}(L/2)$ has a very weak system-size dependence,
whereas $\rho_{\rm s}$ and DW tend to vanish for $L\to \infty$.
}
\label{HISF}
\end{figure}
%%%%%%%%%%%%%%%%%%%%%%%%%%%%%%%%%%

We examined the system-size dependence of the HI and MI phase 
boundaries in Fig.\ref{QMCPD1}, Fig.\ref{QMCPD2}, and Fig.\ref{QMCPD3}. 
Fig.\ref{QMCPD_HI} shows that the phase boundaries obtained 
by the present SSE-QMC simulation does not have a large system size dependence.
In the very vicinity of the tip of the HI phase, we found that the string order gradually 
loses a step-wise behavior, 
and simultaneously the very low SF density starts to appear.
From these behaviors of the order parameters, 
a clear phase boundary was not obtained in the present simulation in the very
vicinity of the tips of the MI and HI.

We also studied the system-size dependence of the HI with $\rho=1$,
which is observed in Fig.\ref{QMCPD3corr} (f).
We plot $G_{\rm string}({L/2})$ as a function of ${1/L}$ in Fig.\ref{HISF}, as 
a ``finite-size scaling" analysis.
It is interesting and also important to see a ``finite-size scaling" of the SF and DW.
See Fig.\ref{HISF}. 
From these results, it is expected that the finite SF and DW in the $\rho=1$ HI is a 
system-size effect.
It should be remarked here that 
while the stochastic Green-function QMC simulation in Ref.\cite{1DDMRG2}
exhibits a strong system size dependence of the string order in the HI, 
the present SSE-QMC simulation does not have such a strong dependence
in the string order.
The difference may stem from the fact that while the Green-function QMC 
is applied to the canonical-ensemble system, our SSE QMC is applied to the
grand-canonical ensemble. 
 
The phase diagrams in Fig.\ref{QMCPD1}, Fig.\ref{QMCPD2} and Fig.\ref{QMCPD3} 
should be compared with the results  of the previous works in which 
the average density is fixed, i.e., the canonical ensemble.
In Refs.\cite{1DDMRG,2DDMRG}, by means of the DMRG,
the $(U-V)$ phase diagram for the $\rho=1$ was obtained.
For the case of $n_c=2$, our results are in good agreement with those in 
Ref.\cite{1DDMRG,2DDMRG}.
However, the SF exists between the $\rho=1$ MI and 
HI as in Fig.\ref{QMCPD1} and Fig.\ref{QMCPD2}, whereas
the SF does not exist there in the phase diagram obtained 
in Refs.\cite{1DDMRG,2DDMRG}.
The phase diagram of $V=4.0$ with $n_c=3$ in Fig.\ref{QMCPD2}
is in good agreement with 
that obtained by DMRG in Refs.\cite{1DDMRG,1DDMRG2,1DDMRG3} for 
$\rho=1$, that is, the phase transitions from the MI, SF, HI, and DW take place
as the value of $U/J$ decreases.
The other parts of the phase diagrams and the calculations of the order
parameters in Fig.\ref{QMCPD1}$\sim$Fig.\ref{QMCPD3corr} are new results.

%%%%%%%%%%%%%%%%%%%%%%%%%%%%%%%%%%%%%%%%%%%%%%%%%%%%%%%%%%%%%%%%
\subsection{Phase diagrams for $V=2.0$ with $n_{c}=4$, $5$ and $6$}
%%%%%%%%%%%%%%%%%%%%%%%%%%%%%%%%%%
%Fig.11
\begin{figure}[t]
\centering
\includegraphics[width=6.5cm]{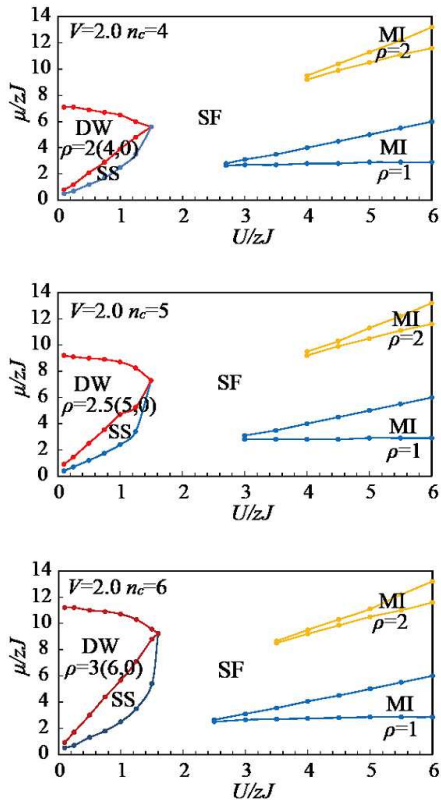}
\caption{(Color online) Phase diagrams for $V=2.0$, $n_{c}=4$, $5$ and $6$.
Higher-filling MI and DW appear with the SS.
This result shows the $n_c$-dependence of the phase diagram.
}
\label{higher_nc}
\end{figure}
%%%%%%%%%%%%%%%%%%%%%%%%%%%%%%%%%%%%

In the previous subsection, we studied the case $n_c=2$ and $3$, and 
obtained the phase diagrams by calculating various order parameters.
In this section, we consider the system with higher $n_c$.
By a simple application of the Holstein-Primakoff transformation 
for the EBHM with the highest-particle number at each site $n_{c}$, 
the EBHM is mapped into a spin $s=n_{c}/2$ model. 
This transformation connecting two models naively implies that
the HI phase appears in the case of an integer $s$, 
i.e., an even integer $n_{c}$ \cite{Haldane}. 
Strictly speaking, however, the EBHM is not mapped to the simple Heisenberg-type
spin model, but to a spin-$s$ model with complicated interactions \cite{Spin_map}. 
Thus, ground-state phase diagram conjectured by the simple correspondence between
the boson and spin models is not necessarily correct.
Moreover, the one-dimensional system has strong quantum fluctuations. 
Thus, the truncation number of the particle in the SSE, $n_c$, may be 
an important ingredient to determine the ground-state phase diagram.
To this end, we perform the SSE-QMC simulation with higher $n_{c}$ in this subsection. 
 
The obtained phase diagrams in the $(U/J-\mu/J)$ plain are shown in Fig.\ref{higher_nc}.
The MIs with the density $\rho=1$ and $2$ are exist in the phase
diagram as in the previous low $n_c$ case.
Their location does not change substantially from the case of $n_c=3$ (and $n_c=2$).
This result is plausible, as the density fluctuations are small in the MIs.
On the other hand for the DW state, ones with the higher average density 
appear in the $n_c=4$, $5$ and $6$ cases, i.e., $\rho=2.0$, $2.5$ and $3$, respectively.
In the $\rho=2.0$, $\rho=2.5$, and $\rho=3$ DWs, 
the state $|\cdots, 4,0,4,0,\cdots\rangle$, 
$|\cdots, 5,0,5,0,\cdots\rangle$, and $|\cdots, 6,0,6,0,\cdots\rangle$ form.
As seen in Fig.\ref{higher_nc}, the SS also forms between the DW and SF.
However, we could not find a HI similarly to the case of $n_c=3$.
For higher $n_c$, particle number at each site can fluctuate rather freely
compared with the case of lower $n_c$, and as a result, the state with particle
number from zero to $n_c$ appears at each site even in the vicinity of the DW.
This may be the reason for the non-existence of the HI.
The above numerical results also indicate that even though 
the system parameters are set around unit filling, 
the EBHM in the grand-canonical ensemble cannot be directly connected to 
the spin-1 model because the HI phase does not exist.
As a future work, to clarify the above problem, other numerical methods, e.g., the DMRG, 
exact diagonalization should be applied to the EBHM of higher $n_c$.

%%%%%%%%%%%%%%%%%%%%%%%%%%%%%%%%%%%%%%%%%%
\section{Discussion and conclusion}

%%%%%%%%%%%%%%%%%%%%%%%%%%%%%%%%%
%Fig.FSS
\begin{figure}[t]
\centering
\includegraphics[width=7cm]{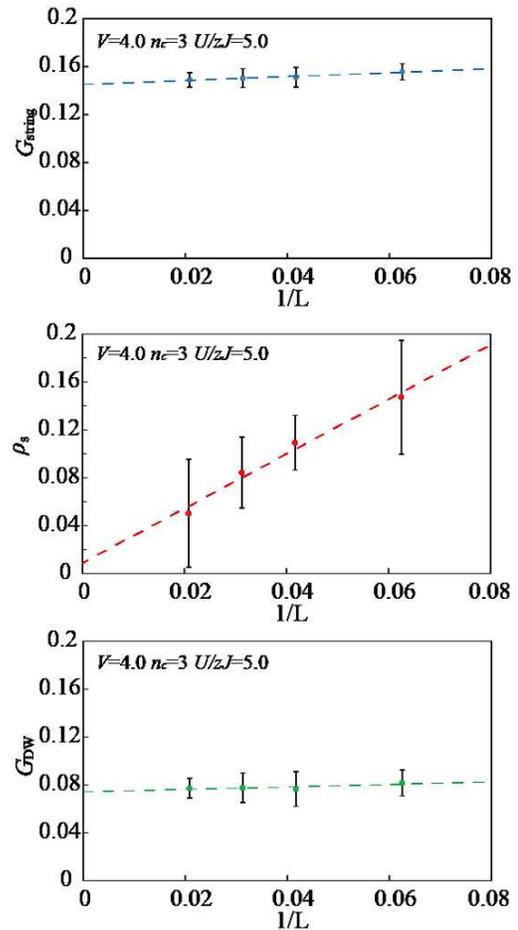}
\caption{(Color online) 
String order (upper panel), SF (middle panel) and DW (bottom panel) in 
the $\rho=1/2$ HI-DW state. SF density tends to vanish for large $L$.
On the other hand, both the string and DW orders have a finite value for large $L$.
However, the string order is larger than the DW order, and this behavior indicates
that unexpected state exists for $\rho=1/2$, i.e., the HI-DW state.
}
\label{HIFSS}
\end{figure}
%%%%%%%%%%%%%%%%%%%%%%%%%%%%%%%%%%%%

In this work, we studied the phase diagrams of the EBHM with the NN repulsion
by means of the SSE-QMC simulations.
We considered the grand-canonical ensemble of the system at low fillings and 
found that the model has a very rich phase diagram.
In the present study, the highest particle number at each site, $n_c$, is a 
controllable parameter as well as the parameters in the Hamiltonian and the filling
factor.
Then, we obtained the phase diagrams for the systems 
without fixing the filling factor.
This is in strong contrast with the previous studies, 
in which the EBHM was investigated in the canonical ensemble 
with specific fillings and also with
a small highest particle number such as $n_c=2$ and 3.
In this sense, we have obtained the global 
and detailed phase diagrams compared to those of the previous works.

In the SSE-QMC simulation, the measurement of the order parameters 
clarifies phase boundaries and clearly exhibits physical properties of each phase,
and the phase diagrams have very small system-size dependence.
Most of the results are in good agreement with the previous works, which study
the EBMH in the canonical ensemble at the unit filling. 
Besides the MI, SF and DW states, there exist the HI and SS.
We also found rather strong $n_c$-dependence of the phase diagram.
This result seems important for the experimental set up to 
observe the phases in the 1D EBHM, in particular, the HI.

%%%%%%%%%%%%%%%%%%%%%%%%%%%%%%%%%
%Fig.05HIsnap
\begin{figure}[t]
\centering
\includegraphics[width=5.5cm]{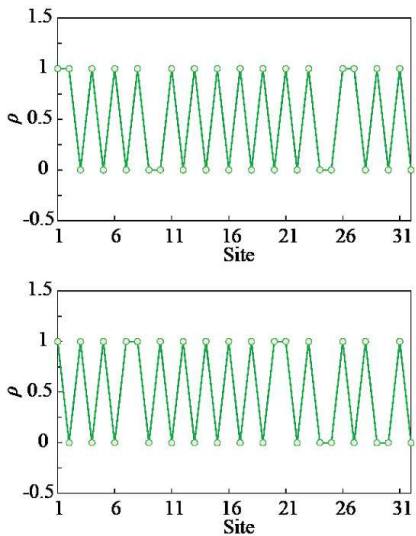}
\caption{(Color online) 
Typical configurations of the $\rho=1/2$ HI-DW
obtained by the SSE-QMC simulation.
In the DW-like background, a local excess of particle is compensated by a pair of
holes.
Distance between them can be considerably large as the MC simulation indicates.
The DW is weakened by these fluctuations.}
\label{HIconfig}
\end{figure}
%%%%%%%%%%%%%%%%%%%%%%%%%%%%%%%%%%%%

We used the string order parameter $G_{\rm string}(L/2)$, defined as in 
Eq.(\ref{OPs}), for searching HIs at various filling factors.
For the system of integer fillings, this quantity is often employed and it
has a real value, whereas for non-integer fillings, $G_{\rm string}(L/2)$ 
can be a complex number.
However, our numerically study reveals that it is always real.
One may wonder what a finite value of $G_{\rm string}(L/2)$ physically means
for fractional fillings.
For the case of the half filling $\rho=1/2$, the numerically observed positive value of 
$G_{\rm string}(L/2)$ indicates that certain specific configurations of the boson
are realized there.
It is also interesting and important to examine the ``finite-size scaling" of
$G_{\rm string}(L/2)$, $\rho_{\rm s}$ and $G_{\rm DW}(L/2)$ for the $\rho=1/2$ state.
The results are shown in Fig.\ref{HIFSS}.
It is obvious that the SF density $\rho_{\rm s}$ tends to vanish for large $L$.
$G_{\rm string}(L/2)$ and $G_{\rm DW}(L/2)$ both keep a finite value for 
large $L$, whereas numerically  $G_{\rm string}(L/2) \sim 2\times G_{\rm DW}(L/2)$.
This strong enhancement of the string order compared with the DW order indicates
that the $\rho=1/2$ state is {\em not} the genuine DW nor the genuine HI, 
and we dare to call it $\rho=1/2$ HI-DW.
(As Figs.\ref{QMCPD2corr} (a) and \ref{QMCPD3corr} (d) show, 
$G_{\rm string}(L/2) \simeq G_{\rm DW}(L/2)$ in the DW.)
Snapshots obtained in the SSE-QMC simulations are shown in 
Fig.\ref{HIconfig}.
In the DW-like background, a local excess of particle is compensated by a pair of
holes, and the DW is weaken by these fluctuations as distances between 
a particle pair and a hole pair can be considerably large.
The HI-DW may connect with the $\rho=1/2$ DW via a phase transition or a crossover as the NN repulsion $V$ increases.

We have recognized that $G_{\rm string}(L/2)$ always has a non-negative real value 
at other fractional fillings.
This seems to indicate that certain chosen configurations are realized in the states
with $G_{\rm string}(L/2)>0$.
This problem is under study.

In recent papers, we pointed out that some parameter regions of the EBHM
are regarded as a candidate for the quantum simulator of a gauge-Higgs model
on a lattice \cite{SForder}.
This observation is quite important as the dynamical properties of the lattice
gauge theory is a very difficult problem and the quantum simulation using 
ultracold atomic gases can study the time evolution of the system.
It is also important to see how exotic states of the EBHM, e.g., the HI,
are understood from the gauge-theoretical point of view.
 
Therefore, let us consider a gauge-theoretical picture of 
the HI phase that exits in the EBHM with small particle density.
As we explained in the previous works \cite{SForder}, the density fluctuation $\delta\rho_a$ plays a role of an electric field in the gauge theory.
Finite $G_{\rm string}(\ell)$ means that
holons and doublons can move rather freely in the sea of the average
particle density but their spatial order is such as 
$(\cdots, \mbox{holon, doublon, holon, doublon},\cdots)$, where
distances between a holon and adjacent doublons 
(and a doublon and adjacent holons) are arbitrary.
In the gauge theoretical language, doublon and holon correspond to
Higgs particle and anti-particle, respectively.
The finite string order lim$_{\ell \to \infty}G_{\rm string}(\ell)\neq 0$
means that particle and anti-particle can separate for a large distance,
but the above mentioned restriction on the mutual configuration must be
satisfied.
From the above observation, one can say that the HI state is a {\em new state} 
of the gauge theory. 
%%%%%%%%%%%%%%%%%%%%%%%%%%%%%%%%%%
%\bigskip

\acknowledgments
Y. K. acknowledges the support of a Grant-in-Aid for JSPS
Fellows (No.JP15J07370). This work was partially supported by Grant-in-Aid
for Scientific Research from Japan Society for the 
Promotion of Science under Grant No.JP26400246.

%\newpage
%%%%%%%%%%%%%%%%%%%%%%%%%%%%%%%%%
 

\begin{thebibliography}{99}
\bibitem{Spin_map}
By the Holstein-Primakoff transformation with spin $S$, 
the bosonic operator is related to spin operators as $S^{+}_{i}=\sqrt{2S-b^{\dagger}_{i}b_{i}}b_{i}$, 
$S^{-}_{i}=b^{\dagger}_{i}\sqrt{2S-b^{\dagger}_{i}b_{i}}$, and $S^{z}_{i}=S-b^{\dagger}_{i}b_{i}$. 
Without taking the large-$S$ limit, the EBHM does not correspond to a
simple Heisenberg model 
because the higher-oder terms of $b_i$ and $b^\dagger_i$ appearing
from the $1/S$-expansion of $S^{+}_{i}$ and $S^{-}_{i}$ 
lead to complicated interaction terms of bosons, which do not correspond to 
the on-site nor nearest-neighbor interaction terms of the EBHM.
\end{thebibliography}
\end{document}